# Ordered interfaces for dual easy axes in liquid crystals

E. Lacaze[1], A. Apicella[2], M. De Santo[2], D. Coursault[1], M. Alba[3], M. Goldmann[1, 4] and R. Barberi[2]

1. Institut des Nano-Sciences de Paris (INSP), UMR-CNRS 7588, Université Pierre et Marie Curie-Paris 6, 140 rue de Lourmel, 75015 PARIS, France
2. Cemif.Cal and Physics Department, University of Calabria, Via P. Bucci, Cubo 33B, I-87036 Rende, Italy
3. Laboratoire Léon Brilloin (LLB), UMR12 CEA-CNRS, CEA-Saclay, F-91191 Gif-sur-Yvette Cedex, France.
4. Université Paris Descartes, 45 rue des Saint Pères, 75006 Paris Cedex, France.

**Abstract**
Using nCB films adsorbed on $MoS_2$ substrates studied by x-ray diffraction, optical microscopy and Scanning Tunneling Microscopy, we demonstrate that ordered interfaces with well-defined orientations of adsorbed dipoles induce planar anchoring locked along the adsorbed dipoles or the alkyl chains, which play the role of easy axes. For two alternating orientations of the adsorbed dipoles or dipoles and alkyl chains, bi-stability of anchoring can be obtained. The results are explained using the introduction of fourth order terms in the phenomenological anchoring potential, leading to the demonstration of first order anchoring transition in these systems. Using this phenomenological anchoring potential, we finally show how the nature of anchoring in presence of dual easy axes (inducing bi-stability or average orientation between the two easy axes) can be related to the microscopical nature of the interface.

**Introduction**
Understanding the interactions between liquid crystal (LC) and a solid substrate is of clear applied interest, the vast majority of LC displays relying on control of interfaces. However this concerns also fundamental problems like wetting phenomena and all phenomena of orientation of soft matter bulk induced by the presence of an interface. In LCs at interfaces, the so-called easy axes correspond to the favoured orientations of the LC director close to the interface. If one easy axis only is defined for one given interface, the bulk director orients along or close to this axis [1]. It is well known that, in anchoring phenomena, two major effects compete to impose the anchoring directions of a liquid crystal, first, the interactions between molecules and the interface, second, the substrate roughness whose role has been analyzed by Berreman [2]. The influence of adsorbed molecular functional groups at the interface is most often dominant with, for example in carbon substrates, a main influence of unsaturated carbon bonds orientation at the interface [3]. In common LC displays, there is one unique easy axis, but modifications of surfaces have allowed for the discovery of promising new anchoring-related properties. For instance, the first anchoring bi-stability has been established on rough surfaces, associated with electric ordo-polarization [4] and the competition between a stabilizing short-range term and a destabilizing long-range term induced by an external field, can induce a continuous variation of anchoring orientation [5]. More recently, surfaces with several easy axes have been studied extensively. It has been shown that control of a continuous variation of director pretilt, obtained in several systems [6, 7], is associated with the presence of two different easy axes, one perpendicular to the substrate (homeotropic) and one planar [7, 8]. Similar models can explain the continuous evolution of anchoring between two planar orientations observed on some crystalline substrates [9]. However, in the same time, two easy axes can also lead to anchoring bi-stability [10, 11] or discontinuous transitions of anchoring [9], which is not compatible with the model established to interpret observed control of pretilt. In order to be able to predict if bi-stability or continuous combination of the two easy axes occurs for one given system, it becomes necessary to understand the microscopic origin of the easy axes.

From an experimental point of view, modern investigation techniques, such as low current Scanning Tunneling Microscopy (STM), allow to obtain well defined informations about the arrangement of liquid crystals molecules at the interface with the substrate. This gives the possibility to relate the characteristics of the anchoring to the presence on the surface of two easy axes and to the angle of disorientation between them for smectic A and nematic phases. In this work we study the role of adsorbed dipoles and alkyl chains for the formation of dual easy axes. This leads us to establish a phenomenological model, as an extension of the well-known Rapini-Papoular one [12], but in direct connection with the microscopically revealed easy axes. This permits an interpretation of our anchoring bi-stability observations, in particular in term of a first order anchoring transition, but also integrates other experimental observations, like the ones of continuous variation of director pretilt.

For such a purpose, we use single crystals of molybdenite as substrates. It is well-known that on crystals, such as graphite or molybdenite ($MoS_2$), the nCB molecules form ordered interfaces, whose molecular structure can be revealed by STM [13, 14]. In reference [11], the bi-stability of anchoring of 8CB adsorbed on $MoS_2$ has been evidenced, by combining STM, X-ray diffraction and optical microscopy. In this system, an ordered interface is created. STM shows that this ordered interface can be described by an alternating orientation of the cyanobiphenyl groups on the substrate, leading to two main orientations for the adsorbed dipolar groups. This corresponds to two easy axes for the planar anchoring, 35° disoriented, finally leading to a bi-stability of anchoring, as probed by X-ray diffraction and optical microscopy. We now extend this work to the systems $11CB/MoS_2$, $10CB/MoS_2$ and $5CB/MoS_2$.

**Experimental**
$MoS_2$ natural single crystals come from Queensland (Australia), supplied by The Ward Company, N.Y. This lamellar compound can be easily cleaved, thereby revealing a clean surface parallel to the basal planes. The surface is composed of sulphur atoms organised in an hexagonal lattice ($a_{MoS2}$ = 3.16 Å as cell parameter), with a mosaicity smaller than 0.02°, as checked by X-ray diffraction. The 10CB and the 5CB come from Merck, Japan; the 11CB has been kindly purchased by Pr. Hara in Riken, Japan. All nCB films have been prepared by melting at 80 °C the organic material on top of a freshly cleaved $MoS_2$ substrate. An organic film, entirely covering the $MoS_2$ surface, is obtained with a thickness varying between 10 and 10000 Å.

In STM experiments, the tip penetrates the non conductive bulk if it is thin and soft enough and probes the structure of the adsorbed molecules. We consequently locate the tips on top of the thinnest LC part, revealing the systematically ordered structure of the nCB monolayer adsorbed on top of the $MoS_2$ substrate. STM experiments have been performed with the low current STM head of a multimode Veeco apparatus (Nanoscope IIIA). Platinum tips, including 20% of iridium for the rigidity, have been used. The measurements have been performed at ambient temperature, but we can assume that the adsorbed structures remain stable within a large temperature range, including part of the isotropic phase, since this is the case for $8CB/MoS_2$ interface up to 120°C [15].

Optical microscopy experiments were performed on a polarising microscope LEICA DMR fitted with a CCD colour camera and a digitizing system for image acquisition. OM images were obtained in the reflection mode, due to the opacity of the $MoS_2$.

X-ray diffraction experiments were performed on the synchrotron beamline CRG-D2AM at the ESRF (Grenoble, France) equipped with 7-circle diffractometer. We used a standard configuration : photon energy at 8 keV, horizontally mounted sample oriented by a goniometrical head in order to explore the whole reciprocal space. The full beam spot was defined by a pair of slits leading to a beam size of 200 x 100 μm close to the sample and the incident intensity was monitored by a diode. The diffracted intensity was scanned parallel to the sample plane by a solid state detector and recorded at the smectic momentum transfer $Q_S$ = 0.18 Å$^{-1}$ of the 11CB molecules. The in-plane and out-of-plane resolutions were of the order of 0.05°. Measurements have been performed with 11CB only, at 30°C in the super-melted smectic phase.

**Results**

We present in the following the results on the systems 11CB/MoS$_2$, 10CB/MoS$_2$ and 5CB/MoS$_2$.

*11CB*

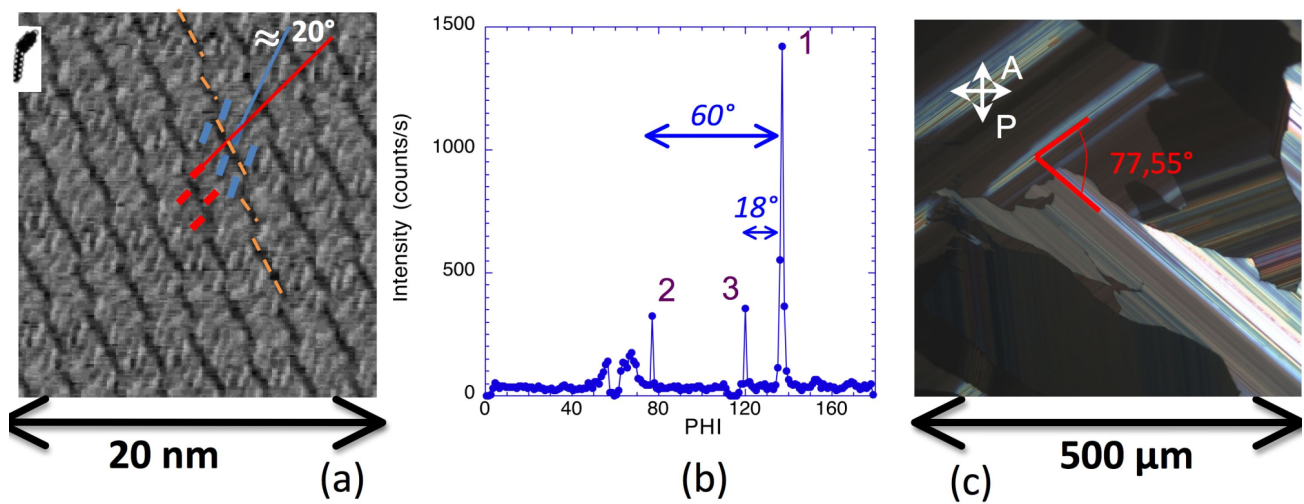

**Figure 1** : a) STM image of the 11CB/MoS$_2$ interface (tunnelling current i$_T$ = 0.37 nA, tunnelling bias V$_T$ = 1.8 V). In the insert, a model of an isolated 11CB molecule is presented. The orientation of the 11CB cyanobiphenyl groups are underlined in red and blue. The [100] MoS$_2$ direction is underlined by an orange dotted line. b) X-rays diffraction intensity with respect to the sample orientation over 180° of a 0.4 µm thick 11CB film on top of MoS$_2$ in super-melted smectic phase (T = 30°C). Three peaks are observed at 137.5° ± 0.5 (labelled 1), 119.5° ± 0.5 (labelled 3) and 77.5°± 0.5 (labelled 2). The pronounced dips at 0°, 58°, and 115° correspond to the furnace tungsten pillars. c) Optical microscopy images between crossed polarisers, of a 11CB film on top of MoS$_2$ in smectic phase (T = 45°C) [16], of thickness ranging from 0.3 µm to 0.5 µm. Disorientation of 77.5° between the domains is underlined in red.

11CB/MoS$_2$ monolayer is perfectly organized [14, 17], similarly to the 8CB/MoS$_2$ interface [18]. Its structure is characterized by the presence of ribbons with a head-to-tail geometry of the molecules within the ribbons (figure 1a). The head-to-tail geometry leads to alternating orientations of the cyanobiphenyl groups on MoS$_2$ within the ribbons. We previously demonstrated that the straight parts of the ribbons in the 11CB/MoS$_2$ monolayer are parallel to the MoS$_2$ [100] direction (orange dotted lines on figure 1a) [17]. The analysis of the 11CB/MoS$_2$ STM images establishes that the two cyanobiphenyl groups are oriented within one ribbon, respectively close to be perpendicular of [100] direction for one group and at roughly 70° of the [100] direction for the other group (repectively red and blue continuous lines on figure 1a).

On the other hand, x-ray measurements performed on 11CB adsorbed on MoS$_2$, evidence that a 11CB smectic film presents defined anchoring directions (figure 1b). Rotating the sample over 180° to detect the orientations of smectic layers (in a circle of diameter approximately 1mm), we observe diffraction peaks associated with the 11CB period, indicating three different anchoring directions of large domains of uniform liquid crystal anchoring, in agreement with optical microscopy observations (figure 1c). Disorientations of 60° between the anchoring directions of these domains of different anchorings are detected (peak 1 and peak 2 on figure 1b). Due to MoS$_2$ hexagonal symmetry, a patchwork of domains disoriented at ±60° is expected in the adsorbed interface.

Consequently, the 60° disorientations previously mentioned evidence the relation between the 11CB/MoS$_2$ interface domains and the smectic anchoring on top. In other words, one given domain of the 11CB/MoS$_2$ interface imposes defined anchoring directions for the smectic layers, as for 8CB/MoS$_2$ [11]. For one given domain, two possible anchoring orientations of the smectic layers, associated with peak 1 and 3 on figure 1c, disoriented by 18°, are revealed by x-rays diffraction. MoS$_2$ Bragg peaks allow to establish the corresponding orientations of the smectic layers, with respect to the [100] MoS$_2$ directions : respectively -1.6° and 16.4 °, corresponding to orientations of the director at 91.6° and 73.6° with respect to the [100] MoS$_2$ directions. They appear to be quasi-parallel to the adsorbed cyanobiphenyl groups orientations (close to 90° and 70°). This shows that 11CB on MoS$_2$ is an interface characterized, for each domain of the adsorbed monolayer, by two orientations of the cyanobiphenyl groups inducing the presence of two easy axes, finally leading to a bi-stability of anchoring for 11CB film, similarly to 8CB on MoS$_2$. These results are confirmed by optical microscopy measurements. Between crossed polarisers, domains of planar uniaxial anchoring are extinguished if the anchoring is parallel or perpendicular to the analyzer. This allows to determine the disorientations between the planar anchoring directions of the domains within liquid crystal films with a 90° degeneracy. This degeneracy can be overcome in smectic phase, since defects associated with an homeotropic anchoring at the air interface appear in the liquid crystal film, leading to linear textures perpendicular to the planar uniaxial anchoring direction (figure 1c) [19]. Measurements of the textures consequently directly evidence the uniaxial planar anchoring directions within each domain. Disorientations between the domains of 17.5°±2° and 17.5°± 60 ± 2° are obtained, in good agreement with x-ray results.

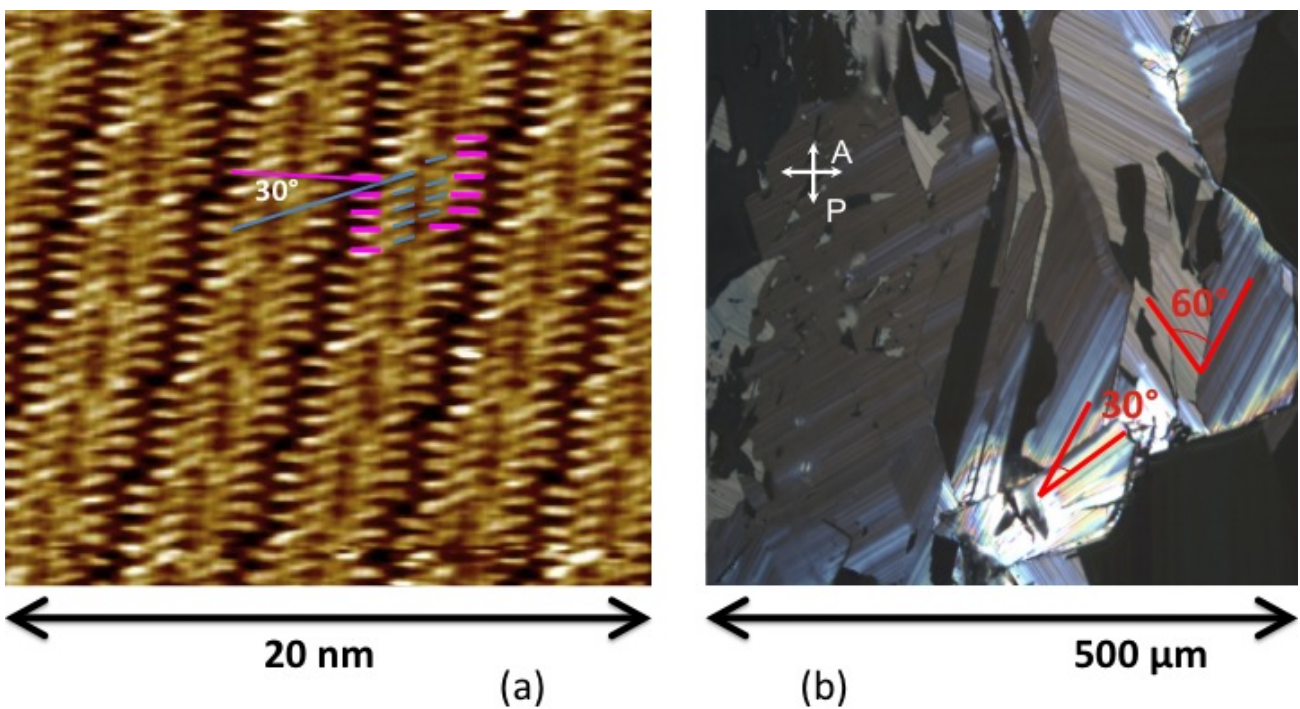

**Figure 2** : a) STM image of the 10CB/MoS$_2$ interface (tunnelling current $i_T$ = 10 pA, tunnelling bias $V_T$ = 1.6 V). The orientation of the 10CB cyanobiphenyl groups are underlined in pink, the one of alkyl chains in blue. b) Optical microscopy images between crossed polarisers, of a 10CB film on top of MoS$_2$ in smectic phase (T=45°C) [20], of thickness ranging from 0.3 μm to 0.5 μm. Disorientations of 30° and 60° between the domains are underlined in red.

Two remarks are necessary at this stage. First, despite a small disorientation between the two easy axes for 11CB (20°), the 11CB liquid crystal director does not choose a unique direction between the two easy axes. Its orientation, along one of the two axes, is associated with anchoring bistability.

Second, in addition to the isotropic phase, 11CB presents only a bulk smectic phase and no nematic one. However the easy axis is imposed by the adsorbed cyanobiphenyl groups orientation, independently of any commensurability between the ordered adsorbed structures and the smectic layers period. The anchoring is thus of nematic type, despite the absence of any bulk nematic phase. This was already the case for 8CB, although 8CB presents both phases, the nematic and the smectic one, depending on the temperature value.

*10CB*

In order to further modify the disorientation of the easy axes at the interface, i.e. the cyanobiphenyl groups orientation, we studied the anchoring of 10CB liquid crystal on the 10CB/MoS$_2$ interface. Similarly to 11CB, in addition to the isotropic phase, 10CB presents a bulk smectic phase only. The interface structure has been previously studied by STM [21] and is displayed on figure 2a. It reveals an overall structure different from the 11CB and 8CB one, characterized by the so-called double-row structure. Each ribbon is associated with two molecules facing each other (and not with only one, of alternating orientation, as for 11CB/MoS$_2$ and 8CB/MoS$_2$). Thus all cyanobiphenyl groups are close to be parallel from each other, since their relative disorientation does not exceed 5° (underlined in pink in figure 2a), as well as the associated alkyl chains (underlined in blue in figure 2a), both being disoriented by 30°. By modifying the bias between tip and sample, the MoS$_2$ substrate can also be probed, revealing that the adsorbed dipoles are again oriented, parallel (±5°) to the [120] MoS$_2$ direction.

When we measure the sample initially measured by STM, using Optical Microscopy, domains are revealed as shown by figure 2b. Each domain is associated with homogeneous planar anchoring perpendicular to the stripes, similar to the 11CB/MoS$_2$ ones. However, unexpectedly, disorientations of multiples of 30° are evidenced between the domains (figure 2b). Comparison between orientations revealed by STM and by Optical microscopy demonstrate that for half of the domains, the bulk director is oriented roughly parallel (±5°) to one of three equivalent [120] MoS$_2$ direction, whereas for the other half it is oriented at 30°. This suggests that for half of the domains, the bulk director is oriented parallel to the adsorbed cyanobiphenyl groups of the interfacial monolayer, whereas for the other half of the domains, it is oriented parallel to the alkyl chains. 10CB/MoS$_2$ thus presents anchoring bi-stability, but with only one easy axis associated with adsorbed cyanobiphenyl groups, whereas the other one appears parallel to the alkyl chains.

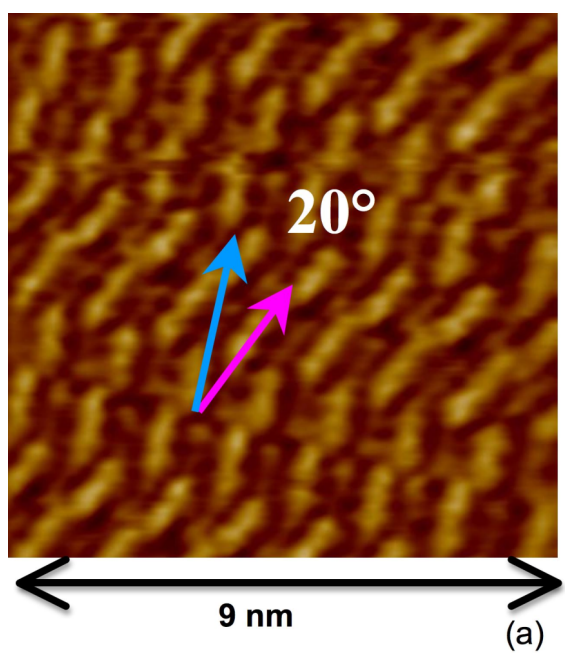

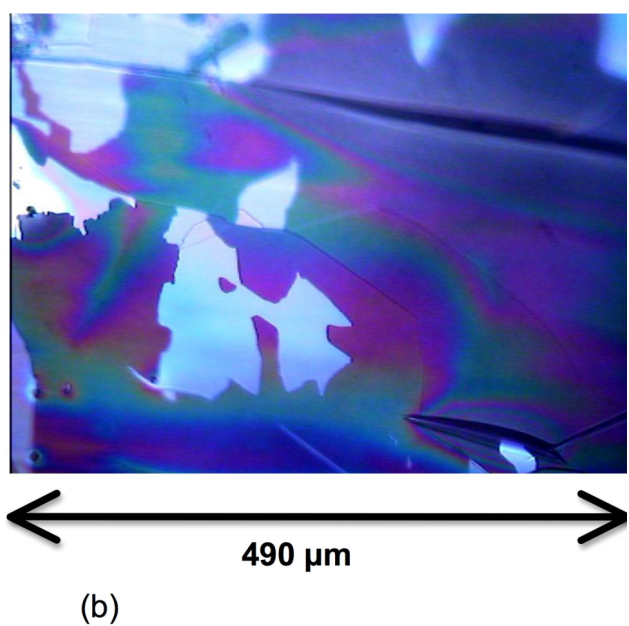

490 μm

9 nm

(a)  (b)

**Figure 3** : a) STM image of the 5CB/MoS$_2$ interface (tunnelling current i$_T$ = 10 pA, tunnelling bias V$_T$ = 1.6 V). The orientation of the 5CB cyanobiphenyl groups are underlined in pink and blue. b) Optical microscopy images between crossed polarisers, of a 5CB film on top of MoS$_2$ in nematic phase, of thickness ranging from 0.2 µm to 0.4 µm. Large domains, uniaxially oriented, are formed in this system.

*5CB*

Finally, we study the 5CB/MoS$_2$ system and its induced bulk anchoring. STM investigations display a single row structure with head to tail geometry of molecules and thus, two different orientations for the adsorbed cyanobiphenyl groups (figure 3a). Ribbons are straight, suggesting that the adsorbed structure is commensurate to MoS$_2$, with ribbons parallel to the [100] direction [17]. Locally the cyanobiphenyl orientations appear close to the ones of the 11CB/MoS$_2$ monolayer. The disorientations of the two adsorbed cyanobiphenyl groups remain small, around 20°, with one orientation closely perpendicular to the ribbons (thus perpendicular to the [100] MoS$_2$ direction) and the other one close to be at 70° of the ribbons. The induced anchoring for 5CB films on top of MoS$_2$ has been studied by Optical Microscopy (figure 3b) and appears different from the 11CB anchoring. It reveals domains considerably larger than the 11CB/MoS$_2$ and 10CB/MoS$_2$ ones, leading to a number of domains per unit of surface at least five times smaller than for 11CB/MoS$_2$ and 10CB/MoS$_2$. No stripes can be evidenced due to the nematic phase of 5CB. Under crossed polarisers, extinctions of the domains are evidenced when the sample is rotated with respect to the analyzer every 30°, in agreement with 60° disorientations between the domains. This result demonstrates that 5CB anchoring is directly imposed by the underlying 5CB/MoS$_2$ interface and by its hexagonal degeneracy, but it also strongly suggests that no bistability occurs for one given domain of the interface, each domain on the contrary imposing a monostable anchoring. In a second step, we established the crystallographic directions of a single crystal MoS$_2$ by x-ray diffraction and compared the anchoring orientations of the nematic film to the substrate crystallographic directions by Optical Microscopy, with a 90° degeneracy due to the measurements between crossed polarisers. It appeared that the nematic domains are extinguished between crossed polarisers when the substrate is rotated by 5° ± 5° (or 35° ± 5°) with respect to the [100] MoS$_2$ direction. Two hypothesis are consistent with such an observation : the nematic director is oriented, either close to the dipolar group perpendicular to the ribbons direction, or along an average direction, between the two adsorbed cyanobiphenyl groups.

**Discussion**

For the 11CB/MoS$_2$ or 8CB/MoS$_2$ systems, the orientation of the two easy axes appear mainly related to the orientation of the adsorbed cyanobiphenyl groups. In nCBs, a strong dipole is located on the cyanobiphenyl group. It is well known that in liquid crystals the nematic director is often oriented parallel to the molecular dipoles [22]. The unsaturated bonds define adsorbed dipoles, which, in turn monitor the bulk director orientation. This is also the case for example of carbon substrates [3]. The simplest case of anchoring corresponds to adsorbed dipoles, all parallel from each other, leading to an also parallel surface nematic director. The bulk nematic director can then be predicted using the Landau de Gennes theory [23], which accounts for the evolution of nematic order parameter from surface to bulk [24, 25]. If the adsorbed dipoles geometry is more complex than a situation in which all dipoles are parallel from each other, as in 11CB/MoS$_2$ or 8CB/MoS$_2$ systems in which several orientations of the cyanobiphenyl groups are defined, our results show that the surface nematic director cannot be considered as the average of the adsorbed dipoles. For a system characterized by two main orientations of the adsorbed dipoles, this would lead to a unique bulk nematic director, oriented between the two adsorbed dipoles. On the contrary, the system appears associated with two easy axes, each axis corresponding to one of the two adsorbed cyanobiphenyl groups orientation. For 10CB/MoS$_2$, only one orientation of the cyanobiphenyl groups is defined whereas two easy axes can be chosen for the bulk director, one associated with adsorbed cyanobiphenyl groups and one at 30°, leading to bi-stability for the system. In fact, in addition to the adsorbed cyanobiphenyl direction, another direction is well defined in one given domain of the interfacial monolayer, the direction of adsorbed alkyl chains. They are indeed expected to be oriented at roughly 150° of the cyanobiphenyl groups, due to, in the same time, the angle between cyanobiphenyl groups and alkyl chains [18] and the tendency of alkyl chains to orient parallel to the [100] direction [21]. Both cyanobiphenyl groups and alkyl chains consequently play the role of easy axes in 10CB/MoS$_2$. Concerning alkyl chains, their tendency to orient bulk director is also well-known. It has been demonstrated in thiol molecules grafted on Au(111) substrates which induce either tilted or homeotropic bulk liquid crystal anchoring [26]. It is finally demonstrated in 10CB/MoS$_2$ system that both cyanobiphenyl groups and alkyl chains similarly play the role of easy axis, most probably due to their similarly well-defined orientation on the substrate. In 11CB, 5CB and 8CB on MoS$_2$, the role of alkyl chains is not obvious. The fact that, in a head to tail geometry, alkyl chains are close to be parallel to the cyanobiphenyl groups (at ±10°) [18] suggests anyway that they could participate to the definition of the easy axes.

Both 10CB and 11CB only present a smectic phase. However they appear oriented by easy axes directions, independently of the fitting of the smectic layers with respect to the interface. This result confirms that the smectic phase anchoring in planar geometry is driven by the same parameters that the nematic one, as already underlined for 8CB/MoS$_2$ [11] or 7BPI on Teflon [27].

It has been demonstrated for 8CB/ MoS$_2$, that two different preparation methods, drop melting and spin coating followed by annealing, lead to the same anchoring textures. Moreover, when bi-stability of anchoring occurs (11CB, 10CB and 8CB), the same number of domains for each orientation of the easy axes, is usually obtained for one given sample. This shows that each anchoring direction corresponds to at least one local equilibrium for the system, once the interface is formed. When a surface is characterized by one easy axis only, the anchoring energy is usually described by a phenomenological expression, the so-called Rapini-Papoular energy (RP) [12], a second order energy, function of the nematic director orientation, θ, with respect to the easy axis, $θ_o$ : $E = W_o \sin^2(θ-θ_o)$. It appeared perfectly relevant for example for very weak anchoring on planar orienting photopolymers [28]. However, in the case of two easy axes, this potential cannot account for two energy minima. It has been shown that an anchoring potential including fourth order terms is necessary to describe complex phenomena like anchoring transitions [9] or bi-stability phenomena [29, 30]. Bi-stability phenomena have been evidenced in tilted anchoring geometries and interpreted using fourth-order terms in an anchoring potential, function of the zenithal anchoring orientation [29, 30]. We now evidence that bi-stability also occurs in planar geometry, associated with two different orientations for the adsorbed dipoles or for the adsorbed dipoles and alkyl chains, which lead to the presence of two easy axes. We thus consider the potential $F_1 + F_2$ associated with each axis, θ being the azimuthal nematic director orientation (planar geometry), with $F_i = W\sin^2(θ-θ_i) + W'\sin^4(θ-θ_i)$. We solved the energy minimization numerically. It appears that W<0 with W/W'=-0.225 leads to two energy minima for F(θ) + F(θ−20), disoriented by 18°, in agreement with the 11CB/MoS$_2$ system whose interface is associated with two easy axes at 20° from each other. W<0 with W/W' = -0.5 leads to two energy minima for F(θ) + F(θ−30), disoriented by 30°, associated with easy axes at 30°, in agreement with the 10CB/MoS$_2$. Finally, W<0 with W/W'=-0.576 describes 8CB/MoS$_2$, with two energy minima, disoriented by 35° for F(θ) + F(θ−32) and two easy axes disoriented by 32° [11, 18].

The difference of coefficients (W, W') between 11CB, 10CB and 8CB, ranging from W/W' = -0.225 to W/W' = -0.576 appears directly related to the exact orientation of the adsorbed dipoles or alkyl chains in the underlying interface, of disorientation ranging from 20° to 32°. The adsorbed interface being extremely stable if the temperature is varied [15], we do not expect any significant variation of the coefficients with the temperature. It would indeed move away the orientation of the energy minima from the one imposed by the dipoles or the alkyl chains adsorbed on the substrate which has never been observed experimentally up to now. The difference between 11CB/MoS$_2$ and 5CB/MoS$_2$, is less obvious. For these systems, the interfaces are similar, associated with two kinds of adsorbed dipoles, disoriented at 20° one from each other. 11CB/MoS$_2$ displays a bi-stable anchoring and 5CB/MoS$_2$, a mono-stable one. This could suggest that, from 11CB/MoS$_2$ to 5CB/MoS$_2$, the W/W' has been varied from a negative W value for 11CB, associated with two energy minima to a positive W value, associated with only one energy minimum. In other words, the evolution from 11CB/MoS$_2$ to 5CB/MoS$_2$ would correspond to a second order transition, possibly associated with the difference in nematic order parameters of smectic 11CB and nematic 5CB phases [31]. This would imply that a continuous evolution of the temperature in these systems, leading to a variation of nematic order, as in 8CB, should induce a continuous disorientation of the nematic order parameter which consequently would not be locked along the orientation imposed by the underlying dipoles, in contradiction with experimental observations. More precisely, if W/W' coefficient of 8CB/MoS$_2$, which presents a variation of 0.1 in order parameter between its two phases, smectic and nematic [31], is only varied by 0.127, half the value necessary to transform the 11CB coefficients into positive values, we obtain W/W'=-0.449, which appears associated with only one energy minimum. This corresponds to mono-stable anchoring, in contradiction with experimental observations, which display similar bi-stable anchoring for both 8CB phases, smectic and nematic [11]. No large variation of (W, W') is thus expected between 11CB/MoS$_2$ and 5CB/MoS$_2$, leading to a second hypothesis: If 11CB/MoS$_2$ and 5CB/MoS$_2$ interface are similar, they differ through the density of adsorbed dipoles and possibly through the two kinds of adsorbed dipoles, which may be not equivalent, which is hardly detected in STM experiments. The non equivalence of the easy axes may be connected to their difference of orientation with respect to the substrate crystallographic directions (one is oriented along the MoS$_2$ [120] direction, whereas the other one is at 20°), inducing difference of interactions with the substrate [18]. This difference of interaction may lead to higher disorder for one of the two dipoles. This disorder would be less significant for 11CB due to the long alkyl chains and the corresponding intermolecular or molecule/substrate interactions, finally leading to a difference in anchoring potential for the two kinds of dipoles of 5CB only, whereas the weights associated with each easy axis remain similar for 11CB or 8CB. This difference of anchoring potential for the two easy axes favours one of the two easy axes, leading to only one stable energy minimum, associated with mono-stable anchoring, oriented close to one of the two adsorbed dipoles. Results of optical microscopy suggest that this easy axis corresponds to the director orientation perpendicular to the [100] direction. The variation from 11CB/MoS$_2$ to 5CB/MoS$_2$ may consequently correspond to a first order transition.

Two types of anchoring potentials can finally be defined in presence of dual easy axes, either W<0 or W>0, leading to different influences for the two easy axes: if W<0, anchoring bi-stability can be obtained and first order anchoring transitions are observed when the weight between the two easy axes is varied; if W>0, mono-stable anchoring is obtained with second order anchoring transitions when the weight between the two easy axes is varied. We demonstrate with nCB series on MoS$_2$ that the first type of anchoring potential is associated with interfaces of two easy axes, intimately mixed, at the molecular level. On the other hand, if one easy axis only is defined, W must be positive to insure one energy minimum only. It is known that microstructured substrates formed by alternating domains of different homogeneous planar anchoring, induce an average planar orientation if the domains are small enough (smaller than 1μm) [33]. This suggests that microscopical structure of the easy axes monitors the nature of the anchoring potential. The second type of anchoring potential would correspond to interfaces formed by domains of homogeneous easy axes, small enough to allow a mixture of both axes at the macroscopical level, the typical size of the domains monitoring the corresponding weights in the anchoring potential. This second type of anchoring potential correctly describes polyimide substrates rubbed by AFM, displaying two easy axes, of zenithal orientation disoriented by 90° [6, 7] or gypsum crystals, leading to planar easy axes disoriented by 40°, which display a second order transition when humidity is varied [9].

This suggests that in these last two systems, small domains of homogeneous easy axes may be formed at the interface. On the other hand, mica substrates displaying a first order anchoring transitions, when humidity is varied [6] may be associated with the first kind of intimately mixed interfacial structure. It is also worthwhile to observe that for ordered interfaces with two easy axes intimately mixed, such as 5, 8, 10 and 11CB on $MoS_2$, the W/W' ratio appears associated with the largest value of W' ever measured with respect to W [29], W' being close to 4 times larger than W in 11CB system. This indeed appears as a necessary condition to obtain bi-stability of two easy axes disoriented by only 20°. Such high values may be related to the high level of ordering of the interface which may also promote high absolute values of anchoring energy, in agreement with the hypothesis that 5CB easy axis with disorder presents a smaller anchoring energy and in agreement also with the results on anchoring energies of smectic phases on crystalline substrates, $MoS_2$ [34] and mica [35].

**Conclusion**
In conclusion, combining x-ray diffraction, optical microscopy and Scanning Tunneling Microscopy, we demonstrate that most of the nCB/$MoS_2$ systems define two easy axes, through the presence of two orientations of adsorbed dipoles or two orientations for adsorbed dipoles and alkyl chains, within the ordered adsorbed interface. These easy axes can induce bi-stability of anchoring which is interpreted using a phenomenological anchoring potential up to the fourth order in sinus of the azimuthal disorientation with negativ coefficients for the second order. We demonstrate that the bi-stable/mono-stable anchoring transition between 11CB and 5CB, associated with similar interfaces on $MoS_2$ corresponds to a first order transition. More generally, we show how the sign of the coefficients in the anchoring potential may be related to the microscopical structure of the adsorbed interface which opens up the route for experimental strategies to control complex anchoring phenomena like control of director pre-tilt or anchoring bistability.

**Aknowledgment**